\documentclass[prl, superscriptaddress,twocolumn]{revtex4}% Physical Review B
\usepackage{epsfig,amsmath,amssymb,latexsym}
\usepackage{graphicx}% Include figure files
\usepackage{dcolumn}% Align table columns on decimal point
\usepackage{bm}% bold math
\usepackage{subfigure}
\begin{document}

\preprint{APS/123-QED}

\title{Electronic Transport Properties of Carbon NanoBuds}

\author{J.~A. F\"{u}rst}
\affiliation{Department of Micro and Nanotechnology, DTU Nanotech,
Technical University of Denmark, DK-2800 Kongens Lyngby, Denmark}
\email{Joachim.Fuerst@nanotech.dtu.dk}

\author{J. Hashemi}
\affiliation{Department of Applied Physics, Helsinki University of Technology, P.O.Box 1100, FI-02015 TKK, Finland}

\author{T. Markussen}
\affiliation{Department of Micro and Nanotechnology, DTU Nanotech,
Technical University of Denmark, DK-2800 Kongens Lyngby, Denmark}

\author{M. Brandbyge}
\affiliation{Department of Micro and Nanotechnology, DTU Nanotech,
Technical University of Denmark, DK-2800 Kongens Lyngby, Denmark}

\author{A.~P. Jauho}
\affiliation{Department of Applied Physics, Helsinki University of
Technology, P.O.Box 1100, FI-02015 TKK, Finland}
\affiliation{Department of Micro and Nanotechnology, DTU Nanotech,
Technical University of Denmark, DK-2800 Kongens Lyngby, Denmark}

\author{R.~M. Nieminen}
\affiliation{Department of Applied Physics, Helsinki University of Technology, P.O.Box 1100, FI-02015 TKK, Finland}

\date{\today}% It is always \today, today,

\begin{abstract}
Fullerene functionalized carbon nanotubes -- NanoBuds -- form a
novel class of hybrid carbon materials, which possesses many
advantageous properties as compared to the pristine components.
Here, we report a theoretical study of the electronic transport
properties of these compounds.  We use both {\it ab initio}
techniques and tight-binding calculations to illustrate these
materials' transmission properties, and give physical arguments to
interpret the numerical results. Specifically, above the Fermi
energy we find a strong reduction of electron transmission due to
localized states in certain regions of the structure while below the
Fermi energy all considered structures exhibit a high-transmission
energy band with a geometry dependent width.

\end{abstract}

\pacs{Valid PACS appear here}% PACS, the Physics and Astronomy
\maketitle

%\section{\label{sec:level1}First-level heading:\protect\\ The line
%%%%%%%%%%%%%%%%%%%%%%%%%%%%%%%%%
%%TEXT STARTS HERE
%%%%%%%%%%%%%%%%%%%%%%%%%%%%%%%%%
Carbon nanotubes (CNT) are among the main candidates for post-CMOS
nanoelectronic devices because of their high carrier mobility as
well as their structural stability.  The electronic, optical, and
transport properties of CNTs depend strongly on their geometry,
offering great versatility, but at the same time posing a huge
challenge because of difficulties in growing and isolating CNT's of
a predetermined type. Defects, impurities and imperfections, as an
inevitable but not necessarily an unfavorable feature of a
real-world nanotube, have also attracted intense attention
\cite{graphene-defect}, because they can modify the electronic
properties of nanotubes to some extent, perhaps even in a
controllable way.

Sidewall chemical functionalization of CNTs is already a
well-established branch of research (reviews are available in
Refs.\cite{charlier:677, Small.1.180}, and recent theoretical
progress is reported, e.g,  in Refs.
\cite{lastra:236806,Bezanilla}). A new member to this family was
introduced by the discovery of a hybrid carbon nanostructure, the
carbon NanoBud (CNB) \cite{main-nanobud} consisting of an imperfect
fullerene covalently bonded to a single-wall carbon nanotube
(SWNT). The CNB's open a new way of functionalizing CNTs, in
particular, because of the high reactivity of fullerenes
\cite{Haddon09171993, fullerenereactivity} suggests the possibility
of further fine-tuning this material via chemical modification.

In order to fully assess the future potential of CNBs a thorough
theoretical examination of their properties is necessary. As far as
we are aware, only some initial studies of CNBs' electronic
structure have been reported thus far \cite{meng:033415,acsNano,Zhu2009}. In
any device application the conductive properties are of crucial
importance. In this work, we undertake a detailed study of the
electronic transport properties of CNBs, using  both \textit{ab
initio} and tight-binding calculations.  As we shall show below, the
transport properties of the CNBs display certain generic features.
We also propose some simple concepts to classify these features,
thereby offering general guidelines for future modeling of more
complicated structures.

%%%%%%%%%%%%%%%%%%%%%%%%%%%%%%%%%%%%%%%%%%%%%%%%%%%%%%%%%%
\begin{figure}
\includegraphics[width=8.5cm] {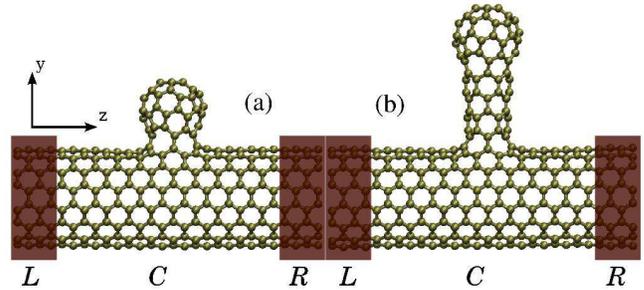} 
\caption{\label{fig1} (Color online) Typical carbon nanobud (CNB) structures
studied in this work.  The CNB consists of an imperfect C60 attached
to an armchair (8,8) single-wall nanotube (SWNT) via a neck region,
made of a (6,0) SWNT. The number of unit cells in the neck region
can vary; panel (a) shows a zero-unit-cell neck (CNB0), while (b)
shows a two-unit-cell neck (CNB2). }
\end{figure}
%%%%%%%%%%%%%%%%%%%%%%%%%%%%%%%%%%%%%%%%%%%%%%%%%%%%%%%%%%%%%%%%%

\par{\it System.
} The experimentally realized CNBs come in a variety of sizes and
shapes \cite{main-nanobud}, and a detailed microscopic knowledge of
the exact atomic positions is not yet known. However, in general the
CNBs can be categorized in two different groups, depending on how
the fullerene is attached to the sidewall of the SWNT
\cite{Nasibulin2007109}. In the first type, a complete fullerene is
covalently bonded to a SWNT via $sp^{3}$-hybridization of carbon
atoms e.g. [2+2] cycloaddition, while in the second type all carbon
atoms are $sp^{2}$-hybridized and the fullerene can be considered as
a part of the SWNT.  In this work we focus on CNBs in the second
group, while the first group will be discussed elsewhere.

Guided by the general features that can be extracted from the
available micrographs, and the density-functional calculations on
the structural stability reported in Ref. \cite{main-nanobud}, we
have chosen to model the CNB structures in the second group as
follows (see Fig. \ref{fig1}). The dome of the CNB is an imperfect
fullerene, C60, with six atoms removed at the apex. The fullerene is
then attached to a (8,8) SWNT via a connecting region ("neck") made
of a varying number of unit cells of a (6,0) SWNT. This
construction allows a relatively smooth joining of the C60 to the
underlying SWNT, even though for the shortest neck regions the
curvature for the connecting bonds is relatively high (see Fig.
\ref{fig1}a). We use the notation CNB$n$, where $n$ is the number of
unit cells of the (6,0) SWNT forming the neck, to describe the structures
studied in this work. As we will show below, the details of the
computed transmission spectra are highly sensitive to the length of
the neck, yet they share certain common features.

%%%%%%%%%%% COMPUTATIONAL %%%%%%%%%%%%%%%%%%%%%%%%%

{\it Computational details.} Once the overall structure of the CNB
has been decided, we relax the system using the Brenner empirical
potential~\cite{Brenner1990} as implemented in the program
\textsc{gulp}~\cite{Gulp}. We have verified that relaxing the
structure in this way leads to insignificant changes in the
transport properties as compared to a high quality density
functional theory (DFT) relaxation. The size of the supercell is chosen so as to provide 10 \AA ~of vacuum between CNBs in neighboring cells which
prevents CNB-CNB interactions. Thus, the supercell size depends on $n$: $22.5\, {\rm\AA}
\times (28+4.26n)\, {\rm\AA} \times 34.46\, {\rm\AA}$. 

The supercell is divided into left (L) and right (R) electrodes
containing 64 C-atoms each, and a central region (C) (see Fig.
\ref{fig1}).

The transport properties were calculated using the nonequilibrium
Green's function method as implemented in the {\sc Transiesta} code
\cite{PhysRevB.65.165401}, which is built based on an atomic orbital
density functional package, SIESTA \cite{0953-8984-14-11-302}. The
GGA PBE functional \cite{PhysRevLett.77.3865} was used to describe
exchange-correlation, and a mesh cut-off value of 100 Ry defining
the real space grid was chosen. A single-$\zeta$ basis set was used
to reduce the computational cost. To benchmark our settings we
relaxed the CNB0 structure and repeated the transmission calculation
using a double-$\zeta$ polarized basis set with all other parameters
unchanged. We found a very good agreement between the results of two
different settings, thereby validating the computationally cheaper
method.

We have also performed tight-binding (TB) based calculations for the
above structures, as well as for larger structures in order to
extract the main trends in the transmission properties. The
electronic Hamiltonian is described with a nearest-neighbor,
orthogonal, $sp^3$ basis set with TB parameters from Charlier {\it
et al.}~\cite{PhysRevB.54.R8377}. Hopping integrals are calculated
within the standard Slater-Koster scheme~\cite{SlaterKoster}. All
the TB calculations are based on relaxed structures as described
above.

However, the
restriction to nearest neighbor hopping allows for a more efficient
recursive Green's function method as described in
Ref.~\cite{Markussen2006}, making the TB calculations very fast. All
our results are calculated in the zero-bias regime.

%%%%%%%%%%%%%%%%%%%%%%%%%%%%%%%%%%%%
\begin{figure}
\includegraphics[width=0.85\columnwidth,angle=-90,viewport= 30 50 600 730,clip] {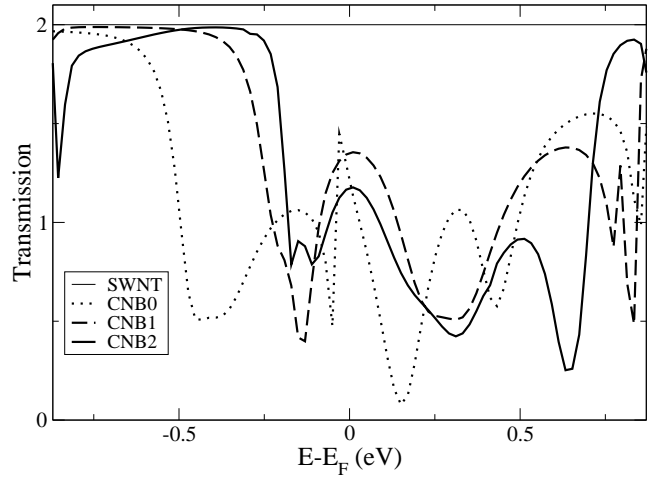} 
\caption{\label{fig2} Transmission as a function of energy for
nanobuds CN0, CN1, CN2, and a pristine (8,8) SWNT. For all neck
sizes the transmission is strongly reduced except for a plateau
region below $E_{F}$.  With increasing neck length this region
shifts upwards in energy and more dips appear at low energies. }
\end{figure}
%%%%%%%%%%%%%%%%%%%%%%%%%%%%%%%%%%%%
%%%%%%%%%%%%%%%%%%%%%%%%%%%%%%%%%%%%%
\begin{figure}
\begin{center}
\subfigure{
\includegraphics[width=0.77\columnwidth, angle=-90,viewport= 60 30 580 730,clip]{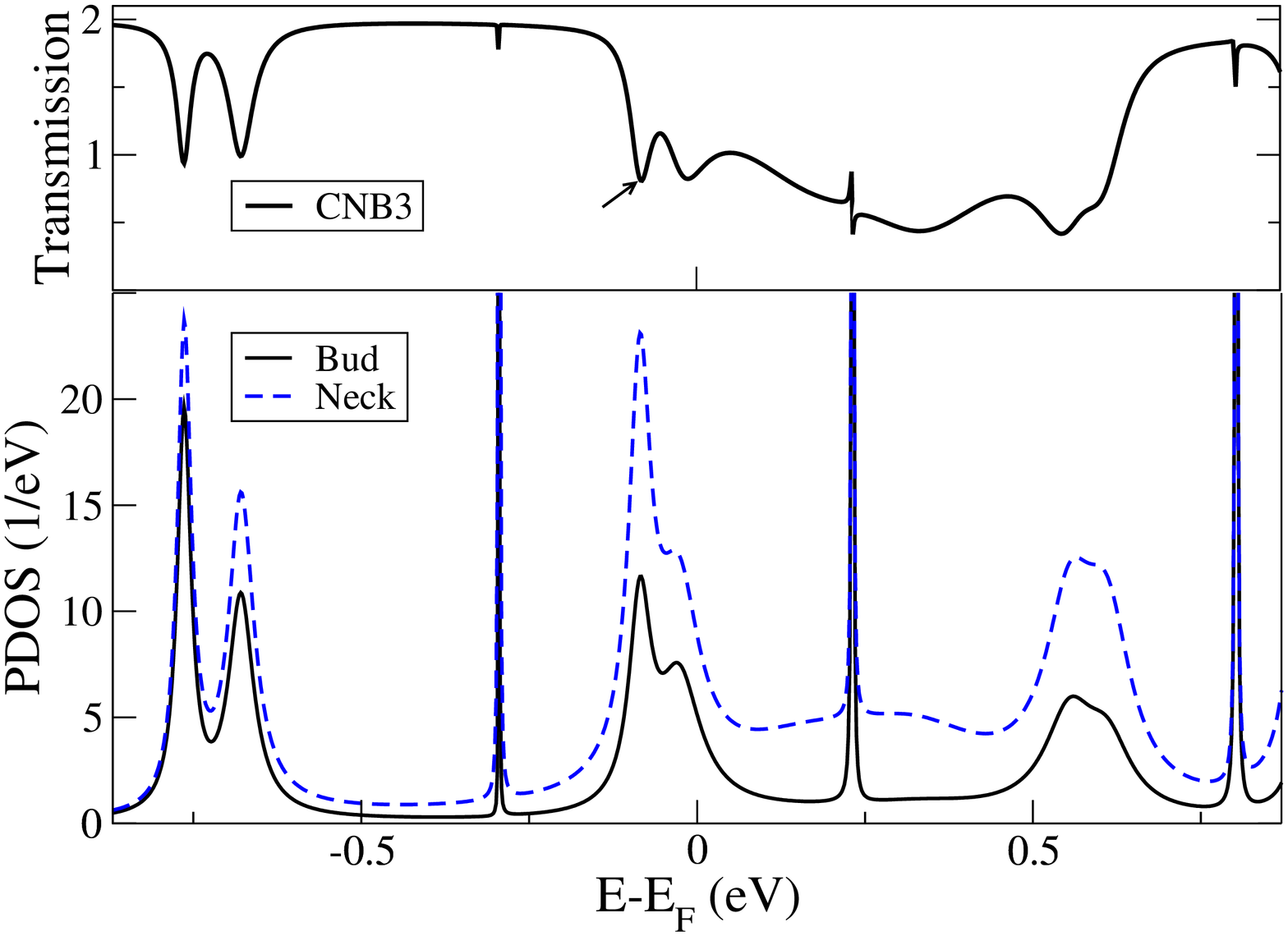}
}
\subfigure{
\includegraphics[width=0.4925\columnwidth, angle=0]{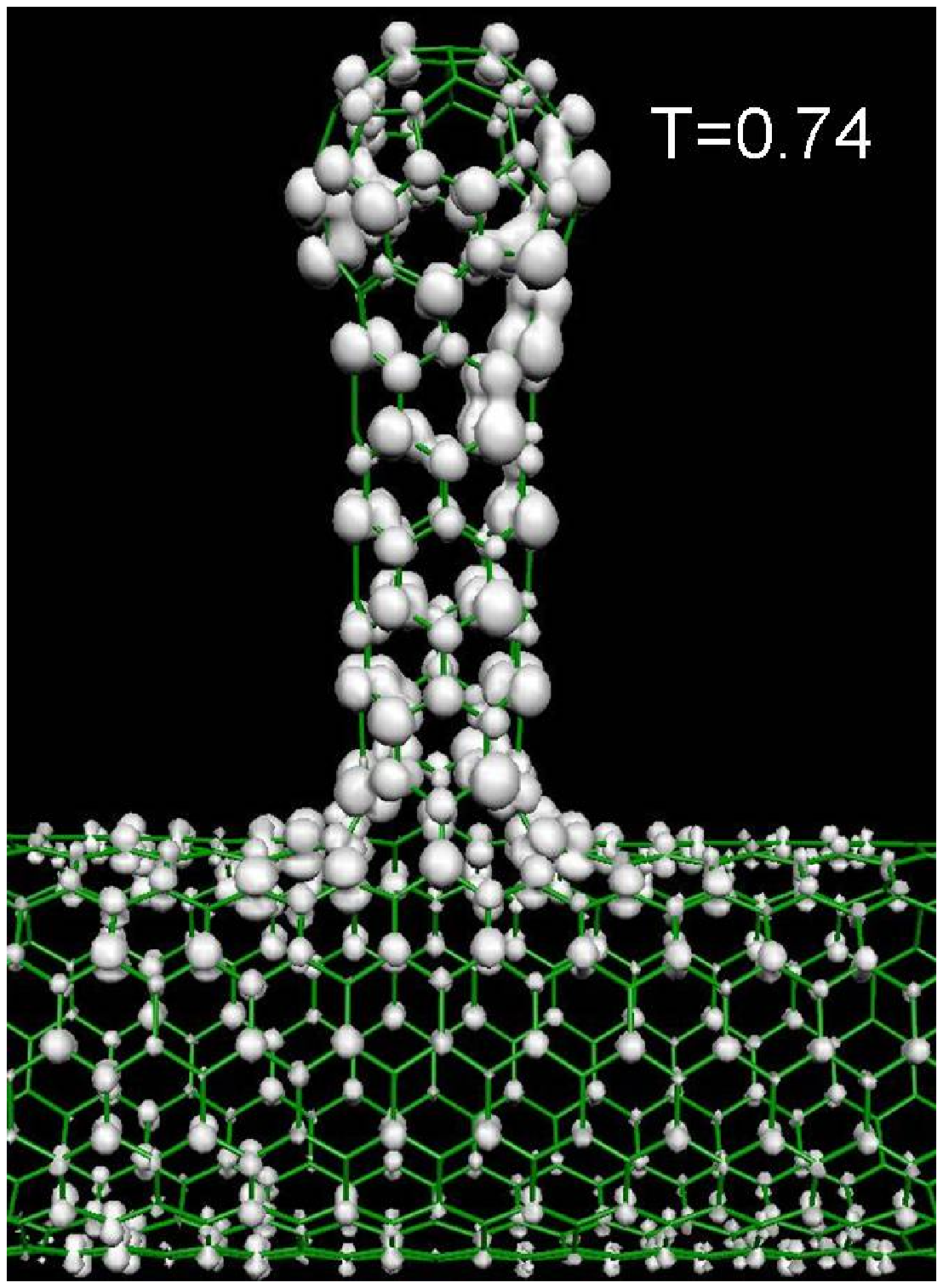}
\includegraphics[width=0.5\columnwidth, angle=0]{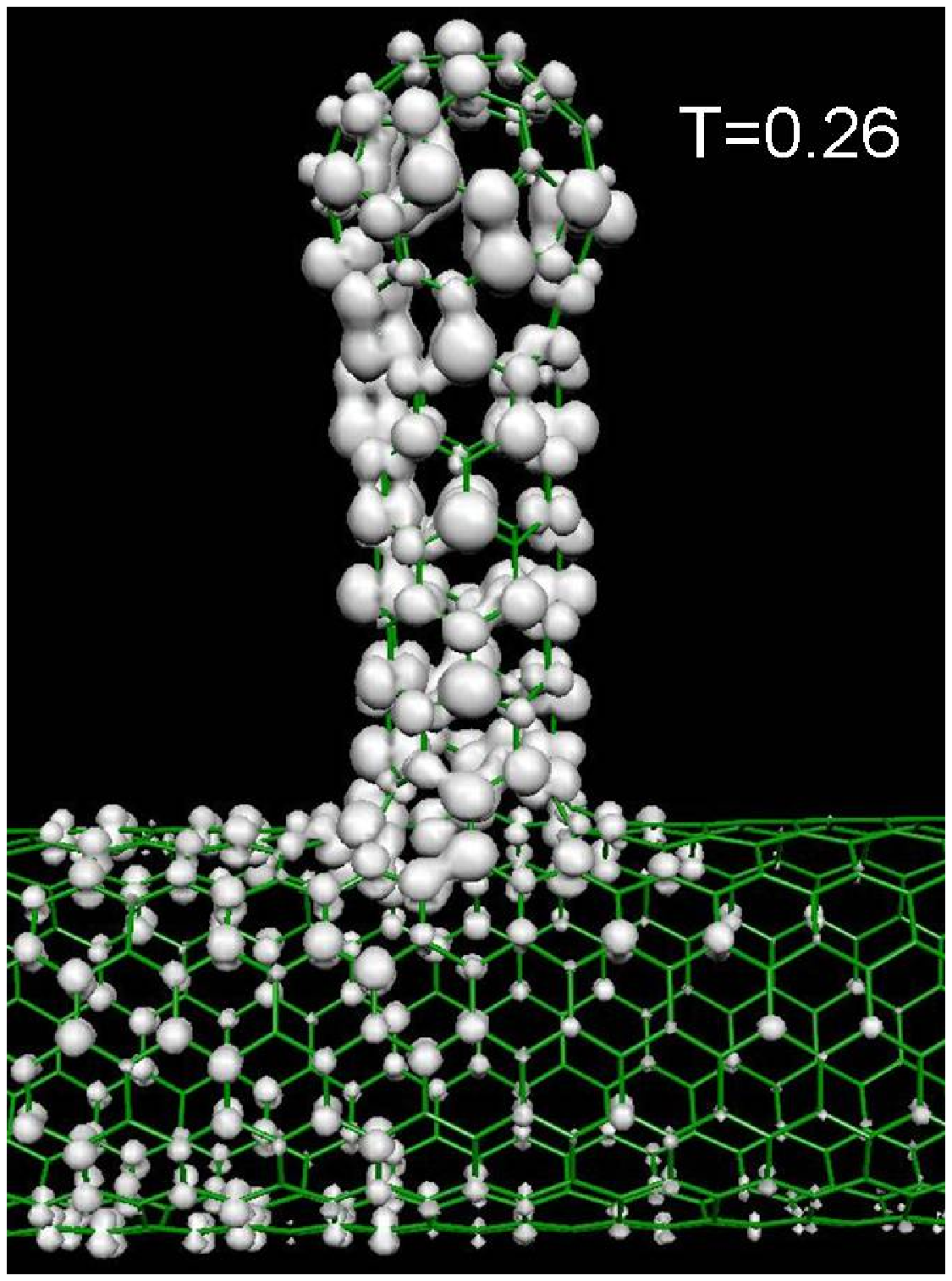}

}
\end{center}
\caption{\label{fig3}(Color online) Projected density of states (PDOS) for the bud
and neck part of the CNB3 system (middle panel). There is a strong
correlation between PDOS and the transmission shown in the top
panel. Bottom: The probability of the eigenchannel scattering states
at the dip in transmission indicated by an arrow in the top panel.
Comparing left ($T=0.74$) and right panel ($T=0.26$) shows that
stronger reduction of transmission is related to stronger
localization of states in the bud and neck.}
\end{figure}
%%%%%%%%%%%%%%%%%%%%%%%%%%%%%%%%%%%%%%%%%%%%%%

%%%%%%%%%%%%%%%% RESULTS %%%%%%%%%%%%%%%%%%%%%

{\it  Results.} Calculated transmissions for three different CNBs
are presented in Fig. \ref{fig2}. An immediate observation is that
in all cases the conductance is reduced at the Fermi energy and
above it. On the other hand, the transmission spectra show a plateau
region below $E_F$ where the transmission is the same as for the
pristine SWNT. The width of this plateau region, however, depends on
the length of the neck region (see also Fig. \ref{fig3} showing
CNB3).

%%%%%%%%%%%%%%%%%%%%%%%%%%%%%%%%%%%%
\setlength{\unitlength}{0.75mm}
%%%%%%%%%%%%%%%%%%%%%%%%%%%%%%%%%%%%
\begin{figure}
\includegraphics[width=0.85\columnwidth,angle=-90,viewport= 30 40 600 730,clip]{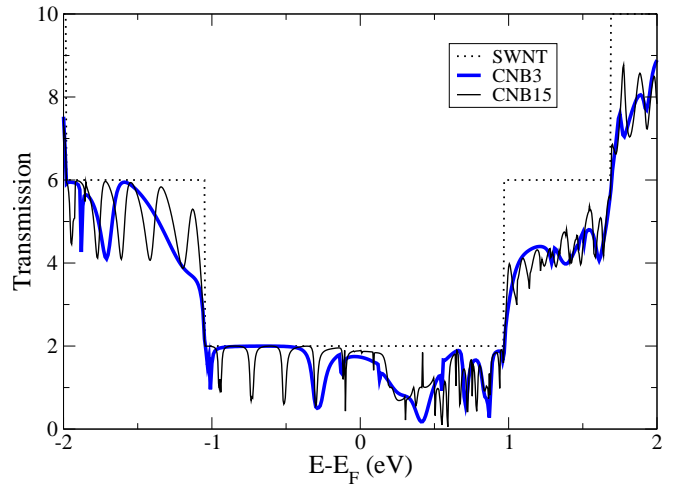}
\caption{\label{fig4}(Color online) The transmission for SWNT, CNB3 and CNB15
calculated with a tight-binding model. The trends from first
principles are well captured for the smaller CNB3 system.
Significantly increasing the system size results in a large number
of dips which below $E_F$ display some degree of periodicity.}
\end{figure}
%%%%%%%%%%%%%%%%%%%%%%%%%%%%%%%%%%%%

Let us next examine the origin for the features in the CNB
transmission in more detail. A drop in the transmission can be
attributed to two sources. First, vacancies are well-known to reduce
the conductance
\cite{vacancy2,vacancy1,PhysRevB.62.7639,JPSJ.68.716}, in particular
when they are adjacent \cite{vacancy1}. The SWNT part of our
structures has 6 vacancies and one therefore expects a strongly
reduced transmission.  Second, localized states in the neck and
bud region should cause strong back-scattering. This can be clearly
seen in Fig. \ref{fig3} (top and middle panel) where we show the
total projected density-of-states (PDOS) of bud and neck region
atoms of CNB3 and the transmission spectrum.  (For comparison, for
one bulk SWNT unit-cell containing 32 C-atoms the main features of
PDOS have values below 1 eV$^{-1}$ and are thus small compared to
neck- and bud-PDOS shown in Fig. \ref{fig3}.) All of the dips in
transmission have a direct correspondence with a peak in PDOS
arising from the states that are localized in the neck and bud
regions. Also, in the plateau part of the transmission the bud and
neck regions have a very low PDOS and the transmission is
essentially the same as for a pristine SWNT. Several features in the
transmission spectrum (e.g., at $E-E_F \simeq$ -0.3, 0.24, or 0.8
eV)  can be interpreted as Fano (anti-) resonances
\cite{Furst2008,Fano1961,NockelStone} between the band states in the
SWNT and the localized states in the bud and neck region.
A similar PDOS-transmission analysis of the region of the SWNT just
below the neck reveals that the localized neck and bud states in
fact extend some 5 \AA ~into the SWNT. Thus, it is not obvious how to connect a particular dip in transmission with a particular
part of the entire region of localization. Nevertheless, the carbon
atoms in this region have a higher reactivity than those located in
the underlying SWNT \cite{Haddon09171993, fullerenereactivity}.
Therefore, the local  electronic structure in the bud/neck regions
can be significantly modified, either by chemical adsorption, or
by adding functionalizing  groups, with a concomitant large change
in conductance. If this tuning can be done in a controllable manner
it makes CNBs a candidate for a sensitive chemical sensor.

The influence of the neck and bud atoms can be further illustrated
by performing an eigenchannel analysis \cite{paulsson:115117} where
the scattering states corresponding to elastic eigenchannels at a
particular energy are extracted. Fig. \ref{fig3}, bottom panel,
shows the scattering states for the two eigenchannels of CNB3 at the
dip in transmission indicated by an arrow in Fig. \ref{fig3}, top
panel. Comparing the spatial distribution of the scattering states
in these two eigenchannels (with transmission probabilities $T=0.74$
and $T=0.24$, respectively), one can clearly see that the more the
scattering states are localized in the neck and bud regions, the
weaker the transmission becomes. Thus, the electron wave function
may extend all the way from the tube to the bud, perpendicular to
the transport direction.  We believe that this feature specific to
CNBs has important implications for field emission.  Specifically,
it is a well-known technological complication that most deposition
techniques for SWNT or graphene based field-emitters result in
sheets lying flat on the substrate \cite{apl-field-emission}.  This
is problematic because field emission occurs preferably from tips,
protrusions, and high curvature features
\cite{PhysRevB.61.9986,PhysRevB.66.241402}. CNBs offer a way of
avoiding this problem, because there will always be buds aligned
with the electric field, and, as our calculations show, the extended
states will supply electrons to the bud and neck states. Indeed, a
significantly enhanced field emission was observed in the original
paper reporting the synthesis of CNBs \cite{main-nanobud}.

Fully \textit{ab initio} calculations for larger CNBs become
computationally very demanding.  We have studied these structures
with the tight-binding (TB) method, which qualitatively reproduces
the main trends of the small CNB \textit{ ab initio} results.
Specifically, the TB-method also leads to a high transmission band
at negative energies (with varying band width), and a
reduced transmission at positive energies, though not as dramatic
as found with the DFT method.  An example is shown in Fig.
\ref{fig3} (top panel, DFT) and Fig. \ref{fig4} (TB). We attribute
this latter discrepancy to  a slightly different description of the
neck and bud states involved in this energy window in the two
methods. Increasing the length of the neck leads to an increase in
the number of dips as seen for the CNB15 system shown in Fig.
\ref{fig4}. These features can be interpreted as standing electron
waves forming in the neck region, in analogy with the analysis
presented by Rubio et al. \cite{RubioPRL1998} for finite CNTs.
Focusing on the energy range -1.0 -- 0.0 eV in Fig. \ref{fig4}, we
identify four dips, with energy separation $\Delta E= 0.21$ eV.
Standing waves in a one-dimensional cavity have an energy separation
$\Delta E=\hbar v_F \pi/L$, where the Fermi velocity $v_F=8.5 \times
10^5$ m/s and $L$ is the cavity length. Equating these two
expressions for the energy separation yields $L=8.4$ nm. This is in
very good agreement with the CNB15 structure of Fig.\ref{fig4}, for
which one can associate the length $L_{\rm CNB}=8.5$ nm (the
combined length for bud and neck is 7.5 nm, and the diameter for the
(8,8) SWNT is 1.0 nm). 
An intriguing feature of all computed transmission spectra is the manifest lack of electron-hole symmetry:
the transmission shows much more structures, i.e., rapid
oscillations, for positive energies while it remains a relatively
smooth function for negative energies. Corresponding rapid oscillations found in the PDOS for the bud/neck region  
may arise due to closer lying energy levels calculated for a bud/neck molecule. We have, however, not been able to prove this statement quantitatively.

We have also performed calculations on CNBs where the underlying
SWNT is semiconducting. Results for (10,0) and (12,0) zig-zag SWNT-based CNBs show, as
in the case of metallic tubes, a strongly reduced transmission
for positive energies, while there is a plateau of high transmission
for negative energies.  These results will be discussed elsewhere.

%%%%%%%%%%%%%%%CONCLUSION%%%%%%%%%%%%%%%%%%
{\it Conclusion.} We calculated the transmission spectrum of carbon
NanoBuds for various geometries. Two common features emerge: the
transmission is significantly reduced at $E_F$ and above it, and
high-transmission bands exist for energies below $E_F$. The electron
wave functions may extend to the neck and the bud as well, and are
likely to have an effect on the field emission properties of CNBs.
The neck region atoms play an important role in the conductance of
the system, and suggest that the conductance can be modified by a
further manipulation of this region.

%%%%%%%%%%%%%%%%%Acknowledgment%%%%%%%%%%%%%%

{\it Acknowledgment.} We gratefully acknowledge numerous discussions
with the authors of \cite{main-nanobud} as well as Martti Puska, Karri Saloriutta and Henry Pinto. This work has been supported by the Academy of Finland through its Center of Excellence Program.  APJ is grateful to the FiDiPro program of the Academy.

%%%%%%%%%%%%%%%%%%%%%%%%%%%%%%%%%%%%
\bibliography{CNB-RC_jof}
 %%%%%%%%%%%%%%%%%%%%%%%%%%%%%%%%%%%%

\end{document}